\documentclass[twocolumn]{emulateapj}
\usepackage{hyperref}
\usepackage[usenames,dvipsnames]{color}
\hypersetup{colorlinks,citecolor=Blue,linkcolor=Red,urlcolor=Blue}
\usepackage[titletoc]{appendix}
\usepackage{amsmath}
\usepackage{natbib}

\begin{document} 

\title{The resilience of \textit{Kepler} systems to stellar obliquity} 

\author{Christopher Spalding$^1$, Noah W. Marx$^{1,2}$ and Konstantin Batygin$^1$} 
\affil{$^1$Division of Geological and Planetary Sciences\\
California Institute of Technology, Pasadena, CA 91125} 
\affil{$^2$Calabasas High School, 22855 Mulholland Hwy, Calabasas, CA 91302}

\begin{abstract}
The \textit{Kepler} mission and its successor \textit{K2} have brought forth a cascade of transiting planets. Many of these planetary systems exhibit multiple members, but a large fraction possess only a single transiting example. This overabundance of singles has lead to the suggestion that up to half of \textit{Kepler} systems might possess significant mutual inclinations between orbits, reducing the transiting number (the so-called ``\textit{Kepler} Dichotomy"). In a recent paper, Spalding \& Batygin (2016) demonstrated that the quadrupole moment arising from a young, oblate star is capable of misaligning the constituent orbits of a close-in planetary system enough to reduce their transit number, provided that the stellar spin axis is sufficiently misaligned with respect to the planetary orbital plane. Moreover, tightly packed planetary systems were shown to be susceptible to becoming destabilized during this process. Here, we investigate the ubiquity of the stellar obliquity-driven instability within systems with a range of multiplicities. We find that most planetary systems analysed, including those possessing only 2 planets, underwent instability for stellar spin periods below $\sim3\,$days and stellar tilts of order 30$^\circ$. Moreover, we are able to place upper limits on the stellar obliquity in systems such as $K2$-38 (obliquity $\lesssim20^\circ$), where other methods of measuring spin-orbit misalignment are not currently available. Given the known parameters of T-Tauri stars, we predict that up to 1/2 of super-Earth mass systems may encounter the instability, in general agreement with the fraction typically proposed to explain the observed abundance of single-transiting systems. 
\end{abstract}

\section{Introduction} 

The ever-growing yield of exoplanetary detections continues to reveal peculiarities between the properties of our own solar system and the galactic norm \citep{Batalha2013,Fabrycky2014}. Among these peculiar features, we highlight two in particular. The first is that our solar system possesses essentially no material closer to the Sun than Mercury, with an orbital radius of 0.4\,AU (e.g., \citealt{Durda2000}). In contrast, extrasolar planetary systems are awash with examples of planets orbiting significantly closer than Mercury \citep{Batalha2013}. 

 A second key aspect of the solar system is that the angular momentum vectors of the eight confirmed planets are mutually inclined by only $\sim1-2\,^\circ$. In the 18th century, this coplanarity inspired the so-called ``Nebular Hypothesis," wherein planetary systems originate from flat (i.e., aspect ratios $\ll$\,unity) disks of gas and dust (\citealt{Kant1755},\,\citealt{Laplace1796}).\footnote{In contrast, the absence of material inside of Mercury's orbit remains mysterious \citep{Batygin2015b}.} Given the ubiquity with which planets form within disks \citep{Hartmann2008}, the expectation is that other planetary systems emerge from their protoplanetary nebula possessing a coplanar architecture. However, the frequency with which this coplanarity is retained over Gyr timescales is not fully understood.
 
 Observational determination of mutual inclinations between extrasolar planetary orbits has proved exceedingly difficult \citep{Winn2015}. Inclined planetary companions are frequently hypothesized as explanations for peculiar signals among transiting planets \citep{Dawson2014a,Lai2017}, and under special circumstances, the orbital properties of these companions may be constrained using transit-timing-variations \citep{Nesvorny2013}. In addition, the variation in transit durations within a given multi-planet system reflect mutual inclinations, but are generally limited to small values owing to the requirement that the planets simultaneously transit \citep{Fabrycky2014}. More loosely, stability arguments have been used to place limits upon mutual inclinations in several systems \citep{Laughlin2002,Veras2004,Nelson2014}.
 
 A separate method of ascertaining mutual inclinations has been to compare the relative numbers of multi-transiting systems to single-transiting systems \citep{Lissauer2011b,Johansen2012,Tremaine2012,Ballard2016}. If, say, planetary systems are typically as coplanar as the solar system, one would expect to observe a larger abundance of multi-transiting systems than from a hypothetical population with larger mutual inclinations. Though conclusions differ in the literature \citep{Tremaine2012}, it is generally difficult to explain the high abundance of single-transiting, relative to multi-transiting systems using a single population of mutually coplanar planetary systems. Rather, some fraction (up to 50\%; \citealt{Johansen2012,Ballard2016}) of systems either possess large mutual inclinations, revealing only one planet at a time in transit, or alternatively this fraction of stars host only one planet. 
 
The aforementioned over-abundance of single systems has been dubbed the ``\textit{Kepler} Dichotomy." The earliest attempt to explain the dichotomy proposed dynamical instability to have generated abundant single systems \citep{Johansen2012}, consistent with the larger typical sizes of single-transiting planets. This idea was disfavoured, largely owing to the unrealistically large masses required in order to trigger planet-planet scattering on Gyr timescales. 

Later work found that mutual gravitational interactions among the planets in a tightly-packed, close-in coplanar configuration are generally unable to excite mutual inclinations of a sufficient magnitude to augment the number of single-transiting systems \citep{Becker2015b}. However, numerous independent investigations have demonstrated that perturbations arising from a massive exterior companion planet are in many cases capable of exciting significant planet-planet inclinations \citep{Lai2017,Becker2017,Mustill2017,Hansen2017}. The presence of an exterior companion may enhance the abundance of single-transiting systems, but requires that these exterior companions are sufficiently close to the inner system of planets to be dynamically-relevant, and possess a sufficiently large mutual inclination to misalign the innermost planetary system.

Recently, \citet{Spalding2016b} demonstrated that an alternative mechanism exists to drive mutual inclinations between close-in planetary systems. Namely, the quadrupole moment arising from a tilted, oblate central star provides a sufficiently large perturbation to misalign the orbits of an initially coplanar planetary system. Moreover, upon simulating the 6-transiting \textit{Kepler}-11 system, the stellar quadrupole was found to drive dynamical instability over a multi-Myr timescale, partly resolving the timescale issue in \citet{Johansen2012}. 
 
The stellar quadrupole-driven hypothesis has the advantage that all planetary systems evolve through an early stage during which the host star is rapidly-rotating and oblate \citep{Bouvier2013}. Furthermore, recent work has demonstrated the feasibility that misalignments between disks and their host stars may be readily excited by gravitational interactions with binary stellar companions \citep{Batygin2012,Spalding2014a,Lai2014}. 
 
The primary goal of this paper is to deduce the ubiquity of the aforementioned instability mechanism across different planetary systems, and to develop insight into the physical mechanism of instability.  In brief, we show that the stellar quadrupole tilts the planetary orbits to a point where the precession rates of their longitudes of pericenter become approximately commensurate. This commensurability drives the eccentricities upward until the orbits cross, triggering instability. 

The paper is ordered as follows. We begin with a description of the numerical model used to simulate the studied planetary systems, including the influence of stellar oblateness. We then discuss our key findings, including the prevalence of instability and the resulting orbital properties. Subsequently, an analytic treatment is presented to provide a physical understanding of the dynamics. After an exploration of potential observational tests, we outline future directions that may lead to greater understanding, and summarise our key conclusions.

\section{Methods}

In order to determine the influence of a tilted, oblate star upon \textit{Kepler} systems in general, we simulated the first 20 million years of a selection of 11 planetary systems. For each system, we performed a suite of 110 $N$-body simulations, where each simulation corresponds to a different combination of stellar obliquity and stellar quadrupole moment. Obliquity is defined as the misalignment angle between the spin axis of the star and the normal vector to the planetary orbits. The quadrupole is defined formally in section 2.2, but essentially captures the rotation-induced equatorial bulge developed by young stars. Throughout each simulation, the stellar quadrupole moment is allowed to decay, reflecting contraction onto the main sequence. For those that remain stable, we compute the mutual inclinations between the remaining planets in order to determine how many of the planets could be observed in transit. 


\subsection{Choice of systems}

Our goal was to determine whether the obliquity-driven instability mechanism proposed in \citet{Spalding2016b} is generic across planetary systems with lower multiplicities than $Kepler$-11. Accordingly, we modelled 6 examples of 2-planet systems, 3 examples of 3-planet systems, and 2 examples of 4-planet systems. We drew the system parameters from real, detected systems where measurements are available of the planetary masses, and where the members are under 25 Earth masses \citep{Jontof-Hutter2016}. The properties of these systems are outlined in Table~\ref{systems}. 

Those systems drawn from measurements in \citet{Jontof-Hutter2016} were deemed insensitive to assumptions made in inferring their TTV masses. Other systems had their masses measured variably from TTV and RV techniques, which largely reflects the range in quoted uncertainties. We use the best-fit masses in our simulations, but include observational uncertainties in Table~\ref{systems}. These uncertainties do not affect our conclusions at a qualitative level, as we briefly discuss below.

Choosing real rather than fabricated systems has two advantages. First, we can be sure that the masses and semi-major axes in our simulations are representative of planetary system architectures known to exist. A second advantage of using real systems is that we may place constraints upon the obliquities of their host stars, given that coplanarity has been retained within the observed systems. Such constraints upon stellar obliquity are particularly valuable in systems of low-mass planets, where alternative techniques for spin-orbit misalignment measurements are notoriously difficult to accomplish \citep{Winn2010,Wang2017}.

\begin{table*}[t]
  \centering
\begin{tabular}{ |p{1.5cm}||p{1.2cm}|p{1.2cm}|p{1cm}|p{1.4cm}|p{1cm}|p{1.4cm}|p{1cm}|p{1.4cm}|p{1cm}|p{1.4cm}|p{0.4cm}|  }
 \hline
 \multicolumn{12}{|c|}{Modelled system parameters} \\
 \hline
Name & $M_\star$\,(M$_{\odot}$) & $R_\star$\,(R$_{\odot}$) & $a_1$\,(AU) & $m_1$\,(M$_{\oplus}$) & $a_2$\,(AU)  & $m_2$(M$_{\oplus}$)& $a_3$\,(AU)  & $m_3$\,(M$_{\oplus}$)& $a_4$\,(AU)  & $m_4$\,(M$_{\oplus}$)&Ref.\\
 \hline
 \textit{K2}-38 & 1.07&1.1 & 0.0505 & $12\pm2.9$ & 0.0965 & $9.9\pm 4.6$&-&-&-&-&(1)\\
 \textit{Kepler}-10 & 0.913& 1.065 & 0.0169& $3.33\pm0.49$ & 0.241 & $17.2\pm1.9$&-&-&-&-&(2)\\
 \textit{Kepler}-29 & 0.979& 0.932 & 0.0922 & $4.5\pm1.5$ & 0.1090  & $4.0\pm1.3$&-&-&-&-&(3)\\
 \textit{Kepler}-36 & 1.071& 1.626 & 0.1153 & $4.45^{+0.33}_{-0.27}$ &  0.1283  & $8.08^{+0.60}_{-0.46}$ &-&-&-&-&(4)\\
  \textit{Kepler}-131& 1.02& 1.03 &  0.12557 & $16.13\pm3.5$ &  0.170752  & $8.25\pm5.9$ &-&-&-&-&(5)\\
   \textit{Kepler}-307 & 0.907& 0.814 & 0.0904 &$7.4\pm0.9$ & 0.105  & $3.6\pm0.7$&-&-&-&-& (3)\\
    \textit{Kepler}-18 & 0.972& 1.108 & 0.0446 &$6.9\pm3.4$ & 0.0751  & $17.3\pm1.9$& 0.117&$16.4\pm1.4$&-&-& (6)\\
   \textit{Kepler}-51 & 1.04& 0.94 & 0.253 &$2.1^{+1.5}_{-0.8}$ & 0.384  & $4.0\pm0.4$&0.509&$7.6\pm1.1$&-&-&(7) \\
   \textit{Kepler}-60 & 1.041& 1.257 & 0.0734 &$4.2\pm0.6$ & 0.0852  & $3.9\pm0.8$&  0.103&$4.2\pm0.8$&-&-&(3) \\
   \textit{Kepler}-79 & 1.17& 1.302 & 0.117 &$10.9^{+7.4}_{-6.0}$ & 0.187  & $5.9^{+1.9}_{-2.3}$& 0.287&$6.0^{+2.1}_{-1.6}$& 0.386&$4.1^{+1.2}_{-1.1}$ & (8)\\
   \textit{Kepler}-223 & 1.13& 1.72& 0.0771 &$7.4^{+1.3}_{-1.1}$ &  0.0934  & $5.1^{1.7}_{-1.1}$ & 0.123&$8.0^{+1.5}_{-1.3}$& 0.148 & $4.8^{+1.4}_{-1.2}$&(9) \\
 \hline
\end{tabular}  
 \caption{The parameters of the simulated \textit{Kepler} systems. Initially, we set all eccentricities to zero. Numbered references (Ref.) are (1)\,\citet{Sinukoff2016} (2)\,\citet{Dumusque2014} (3)\,\citet{Jontof-Hutter2016} (4)\,\citet{Carter2012} (5)\,\citet{Marcy2014} (6)\,\citet{Cochran2011} (7)\,\citet{Masuda2014} (8)\,\citet{Jontof-Hutter2014} (9)\,\citet{Mills2016}.}
  \label{systems}
\end{table*}

\subsection{Numerical Set-up}
We begin by performing numerical simulations of planetary systems orbiting stars with varying degrees of obliquities and quadrupole moments. Throughout, we utilize the mercury6 \textit{N}-body integrator \citep{Chambers1999}, employing the hybrid symplectic/Bulirsch-Stoer algorithm with a timestep of integration set as a fraction 1/20 of the shortest planetary orbital period, which typically conserves energy to better than 1 part in 10$^6$. The planets move under the action of their own mutual gravity, along with that of the host star. Expanded to quadrupole order, the stellar potential energy per unit mass may be written as
\begin{align}
V_{\star}=-\frac{GM_\star}{r}\Bigg[1-\bigg(\frac{R_\star}{r}\bigg)^2J_2\bigg(\frac{3}{2}\cos^2 \theta-\frac{1}{2}\bigg)\Bigg],
\end{align}
where $\theta$ is the angle between the planet's position and the spin axis of the star. The stellar mass and radius are denoted $M_\star$ and $R_\star$, the distance from the center of the star is written $r$, and $G$ is Newton's gravitational constant. The quantity $J_2$ is known as the second gravitational moment and encodes the star's oblateness and internal structure, to quadrupole order. 

\begin{figure}[ht!]
\centering
\includegraphics[trim=0cm 0cm 0cm 0cm, clip=true,width=1\columnwidth]{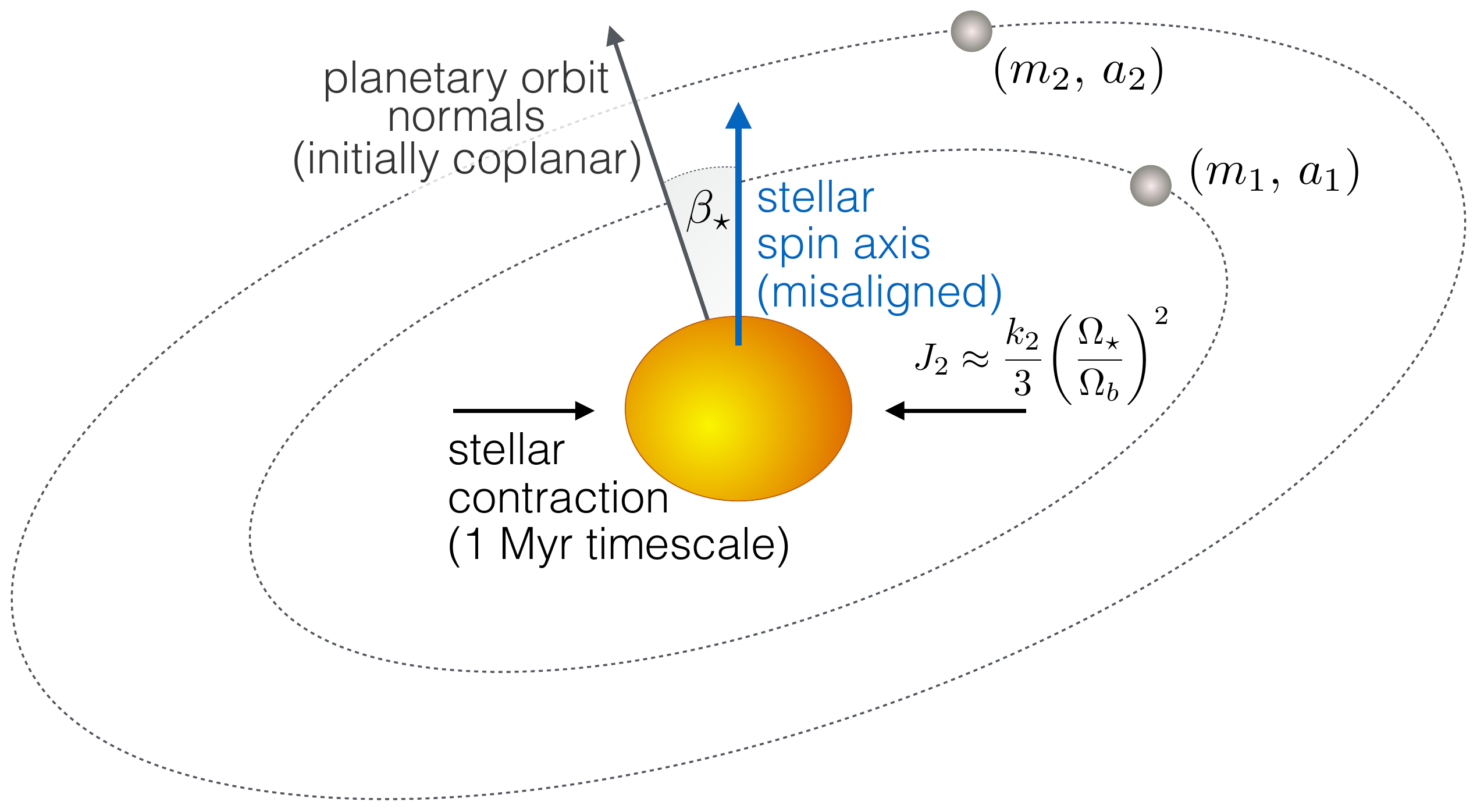}
\caption{A schematic of our numerical simulations. The planetary system is initialized with coplanar orbits, all sharing a mutual inclination of $\beta_\star$ with the stellar spin axis. The star begins with an oblateness parameter $J_2=J_{2,0}$ which decays exponentially on a 1 million year timescale. The simulations are carried out using the symplectic $N$-body integrator mercury6 \citep{Chambers1999}.} 
\label{Schematic}
\end{figure}


A measurement of the gravitational potential around young stars is beyond observational capabilities, and therefore as is a direct measurement of $J_2$. However, it can be shown \citep{Sterne1939,Ward1976} that $J_2$ relates to the observable stellar spin rate $\Omega_\star$ through the expression
\begin{align}\label{J2}
J_2\approx\frac{k_2}{3}\bigg(\frac{\Omega_\star}{\Omega_b}\bigg)^2,
\end{align}
where $k_2$ is the Love number and $\Omega_b\equiv\sqrt{GM_\star/R_\star^3}$ is the stellar break-up angular velocity. Approximation~(\ref{J2}) holds provided that $\Omega_\star\ll\Omega_b$, which is the case for most T-Tauri stars \citep{Bouvier2013}. The benefit of parameterizing $J_2$ as above lies in the ability to directly measure $\Omega_\star$, and to obtain $k_2$ and $\Omega_b$ from stellar models. Specifically, the Love number $k_2$ may be computed from polytropic models of index $\chi=3/2$, yielding $k_2\approx0.28$ \citep{Chandra1939,Batygin2013}. 

Owing primarily to Kelvin-Helmholtz contraction, the product $J_2 R_\star^2$ will decay with time, and with it the quadrupole moment. We choose to parameterize this contraction by supposing that the radius of the star is fixed at $R_\star=2R_\odot$, reflecting the inflated radius typical of young stars \citep{Shu1987}. From this initial state, we allow $J_2$ to undergo exponential decay such that
\begin{align}\label{J2}
J_2(t)=J_{2,0}\exp(-t/\tau_c),
\end{align}
where $J_{2,0}$ is the initial value of $J_2$ and the Kelvin-Helmholtz timescale $\tau_c=1\,$Myr \citep{Batygin2013}.


In our prescription~(\ref{J2}) for $J_2$, we prescribe $M_\star$ and $R_\star$, yielding $\Omega_b$, along with a value $k_2=0.28$. The final component is the stellar spin rate. Here, we must draw from observations of young stars \citep{Bouvier2013} which suggest a distribution of T-Tauri stellar rotation periods ranging from $\sim1-10$\,days. Using the parameters above, we arrive at a range of $J_{2,0}$ given by
\begin{align}
10^{-4}\lesssim J_{2,0}\lesssim 10^{-2}.
\end{align}
Accordingly, in our simulations we choose 11 values of $J_{2,0}$, equally separated in log-space:\footnote{It should be noted that our simulations will begin subsequent to disk-dispersal, meaning that the stellar radius is likely to be somewhat reduced from $2R_\odot$ and so our strongest quadrupole is a slight over-estimate.}
\begin{align}
\log_{10}(J_{2,0})\in\{-4,\,-3.8,\,...\,,-2\}.
\end{align}
\begin{figure*}[ht!]
\centering
\includegraphics[trim=0cm 0cm 0cm 0cm, clip=true,width=1\textwidth]{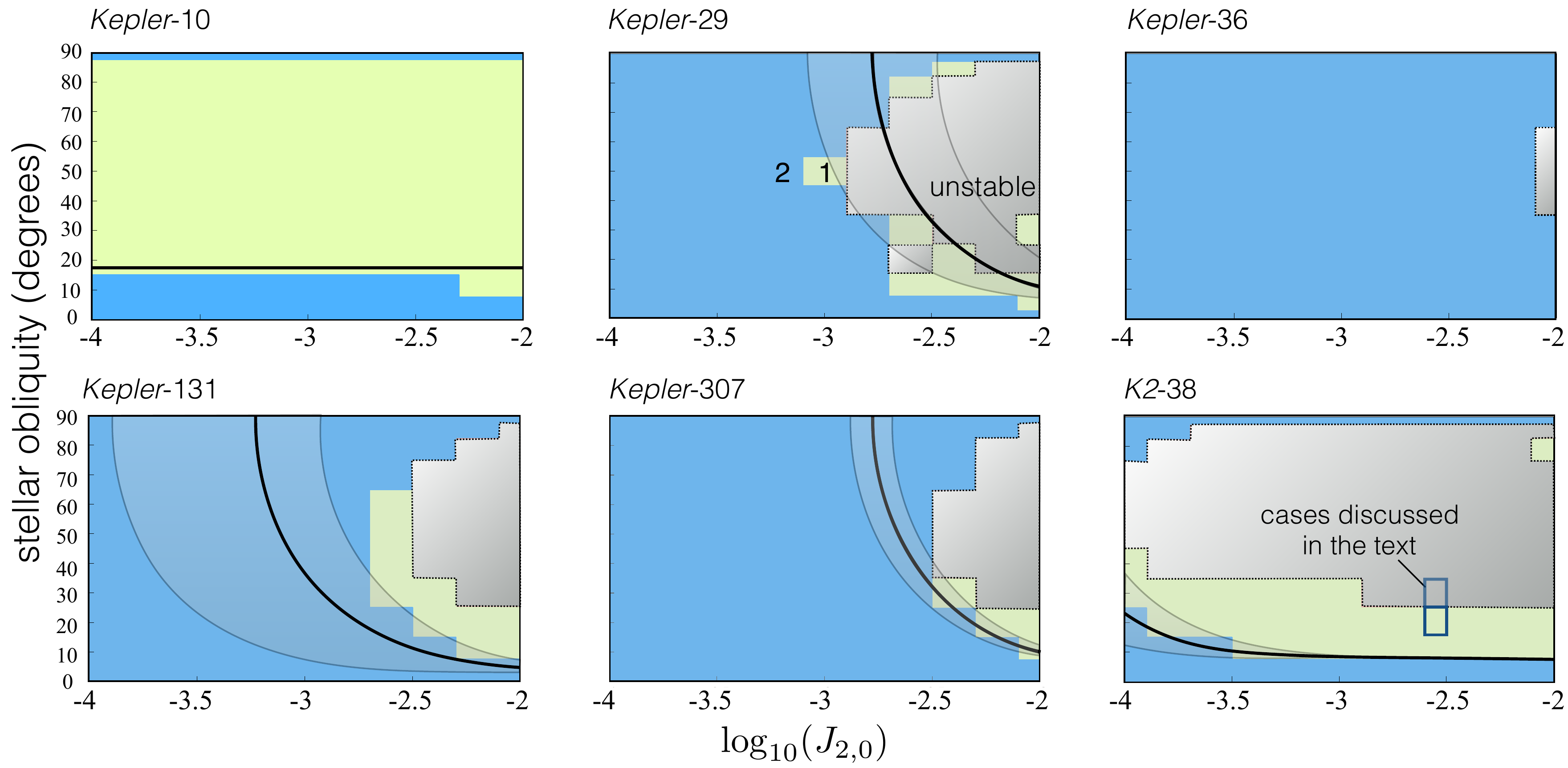}
\caption{The number of planets detectable in transit after 20 million years of simulation from an initially 2-planet configuration. Solid black lines denote the critical obliquity as a function of $J_{2,0}$ predicted to reduce the transit number from 2 to 1 according the the formula \citet{Spalding2016b}. The dotted line outlines the region where one of the two planets was lost owing to dynamical instability. We explore the mechanism of instability in more detail by examining the cases outlined in blue on the plot for $K2$-38.} 
\label{Results_1}
\end{figure*}

 In all simulations, we fix the spin axis of the star to be parallel to the $z$-axis. This approximation is equivalent to the statement that the orbital angular momentum of the planetary system is a small fraction of that contained within the stellar spin angular momentum. The ratio between the orbital to stellar spin angular momenta is given by
\begin{align}
j&\equiv \frac{m_p\sqrt{GM_\star a_p}}{IM_\star R_\star^2 \omega_\star}\nonumber\\
&\approx 0.05\bigg(\frac{m_p}{10 M_{earth}}\bigg)\bigg(\frac{M_\star}{M_\odot}\bigg)^{-1/2}\nonumber\\&\bigg(\frac{P_\star}{10\,\textrm{days}}\bigg)\bigg(\frac{R_\star}{R_\odot}\bigg)^{-2}\bigg(\frac{a_p}{0.1\,\textrm{AU}}\bigg)^{1/2},
\end{align}
where $a_p$ and $m_p$ are the planetary semi-major axis and mass respectively, and $I\approx0.21$ is the dimensionless moment of inertia parameter of the star \citep{Batygin2013}. The smallness of $j$ validates our assumption that the stellar spin axis changes slowly with time and can therefore be approximated as fixed.

We note that the masses we use in our simulations are the best-fit estimates as derived through radial velocity and/or transit timing variations. These techniques lead to substantial uncertainties in the masses of constituent planets (Table~\ref{systems}). Accordingly, our approach here is not necessarily aimed toward a detailed reconstruction of the the history of these systems, but rather, we are using their orbital parameters as general guidelines for ``typical" planetary system parameters.

 As initial conditions, we set all eccentricities to zero, with semi-major axes chosen to fit those measured in the systems today. For each value of $J_{2,0}$, we run simulations with 10 different initial stellar obliquities ($\beta_\star$), spread between 0 and 90 degrees:
\begin{align}
\beta_\star\in\{5,\,10,\,20,\,30,\,40,\,50,\,60,\,70,\,80,\,85\}.
\end{align}
A schematic of the initial set-up is illustrated in Figure~\ref{Schematic}.

\subsection{Determination of transit number}

At uniform time intervals during our model runs, we deduce the maximum number of planets that can be observed transiting at that particular time. Specifically, we compute the mutual inclinations between all of the planetary orbital pairs. Considering a pair of planets $i$ and $j$, the mutual inclination $I_{ij}$ between their orbits is computed using the geometrical relationship:
\begin{align}
\cos(I_{ij})=\cos(I_i)\cos(I_j)+\sin(I_i)\sin(I_j)\cos(\Omega_i-\Omega_j).
\end{align}
Having computed $I_{ij}$, we consider the planets to be removed from a mutually-transiting configuration if the following criterion is satisfied: 
\begin{align}
\big|\sin(I_{ij})\big|\gtrsim \frac{R_\star}{a_i}+\frac{R_\star}{a_j}.
\end{align}

For example, given three planets we compute $I_{12}$, $I_{13}$ and $I_{23}$. If all satisfy the above criterion, the transit number is unity. If $I_{12}$ and/or $I_{23}$ do not satisfy the criterion but $I_{13}$ does, the transit number is 2, etc. Given that the mutual inclination will change with time, potentially bringing the planet pairs into and out of mutual transit, we average the transit number over the final $\sim10^5$ years of the integration.

\subsection{Caveat: Disk Potential}

It is important to point out one confounding factor in our results. We began with an initial condition whereby the planetary system possessed a non-zero inclination with respect to the stellar spin axis. However, in any physical situation like this, it is important to ask how the system was set up in that configuration, especially if that configuration is not a steady-state. Here, the key assumption was that the disk dispersed on a short enough timescale such that the planets inherited the disk's plane exactly. 

To examine this problem, we cannot simply add a disk potential to the numerical simulations, because in that case fixing the stellar precession axis is no longer necessarily valid \citep{Spalding2014a}. The disk will induce a nodal regression upon the planetary orbits of \citep{Hahn2003}
\begin{align}
\nu_{\textrm{disk}}\approx n_p \frac{\pi \sigma a_p^2}{M_\star}\frac{a_p}{h},
\end{align}
where $\sigma$ is the disk's surface density and $h$ is the disk's scale height. We may define the time of disk dispersal as the point at which $\nu_{\textrm{disk}}$ is approximately equal to the nodal regression induced by the stellar quadrupole moment ($\nu_\star=(3/2)J_2(R_\star/a_p)^2$). This criterion corresponds to a disk surface density of 
\begin{align}
\sigma_{\textrm{d}}\approx \bigg(\frac{3 M_\star h J_2}{2\pi a^3}\bigg)\bigg(\frac{R_\star}{a}\bigg)^2\approx 200\textrm{g}\,\textrm{cm}^{-2},
\end{align}
where we used $J_2=10^{-3}$, $h/a=0.05$ and $R_\star=2R_\odot$. 

The surface density of the MMSN at 0.1\,AU is approximately 50,000\,gcm$^{-2}$, meaning that disk dispersal for our purposes happens at the point when the disk possesses roughly $1\%$ of its original mass \citep{Armitage2010}. The final stages of disk dispersal in the inner regions are thought to progress through viscous accretion, subsequent to photoevaporative starvation from gas accreting inwards from the outer disk. The viscous time at $0.1\,AU$ is given by 
\begin{align}
\tau_\nu \approx \frac{a_p^2}{\alpha h^2 \Omega}\approx 200\,\textrm{years} \bigg(\frac{0.01}{\alpha}\bigg)
\end{align}
where $\alpha$ is the Shakura-Sunyaev turbulent diffusivity parameter \citep{Shakura1973,Hartmann2008}. 

The precession timescale arising from the stellar quadrupole, at a similar orbital distance, with $J_2=10^{-2}$ is roughly $200$\,years. Accordingly, the disk dissipates on a comparable timescale to that of stellar-induced precession, and thus the system might reduce its spin-orbit misalignment somewhat during disk dissipation. More work is required in order to investigate this possibility. The timescales and physics governing disk dispersal are poorly understood, and so we leave this aspect of the problem as a caveat, to be returned to once better constraints become available.

\section{Results \& Discussion}

\begin{figure*}[ht!]
\centering
\includegraphics[trim=0cm 0cm 0cm 0cm, clip=true,width=1\textwidth]{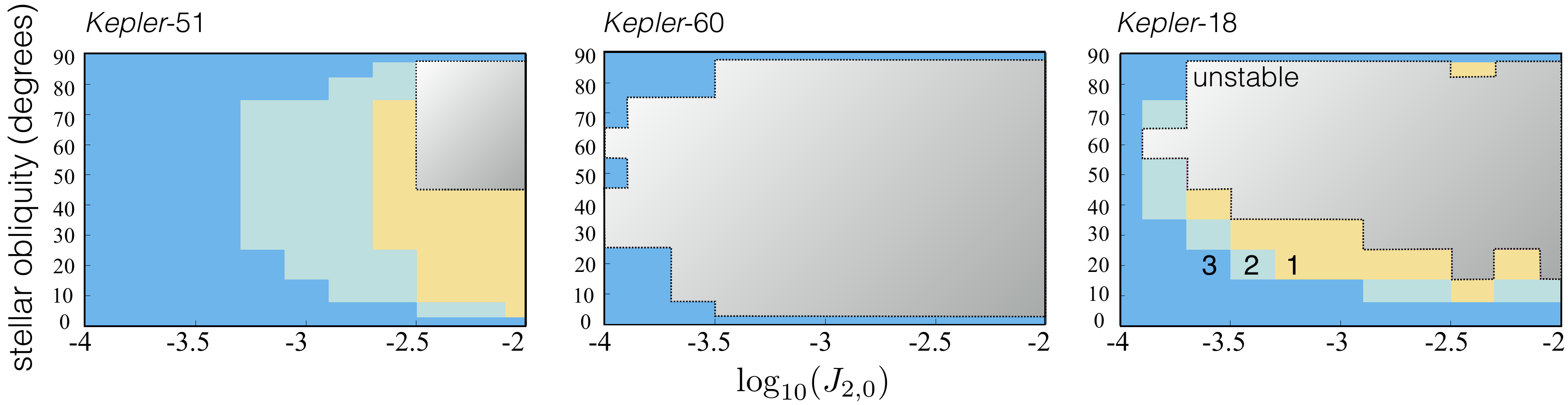}
\caption{The number of planets detectable in transit after 20 million years of simulation from an initially 3-planet configuration. The dotted line outlines the region where one or more planets were lost owing to dynamical instability.} 
\label{Results_2}
\end{figure*}

\begin{figure}[ht!]
\centering
\includegraphics[trim=0cm 0cm 0cm 0cm, clip=true,width=1\columnwidth]{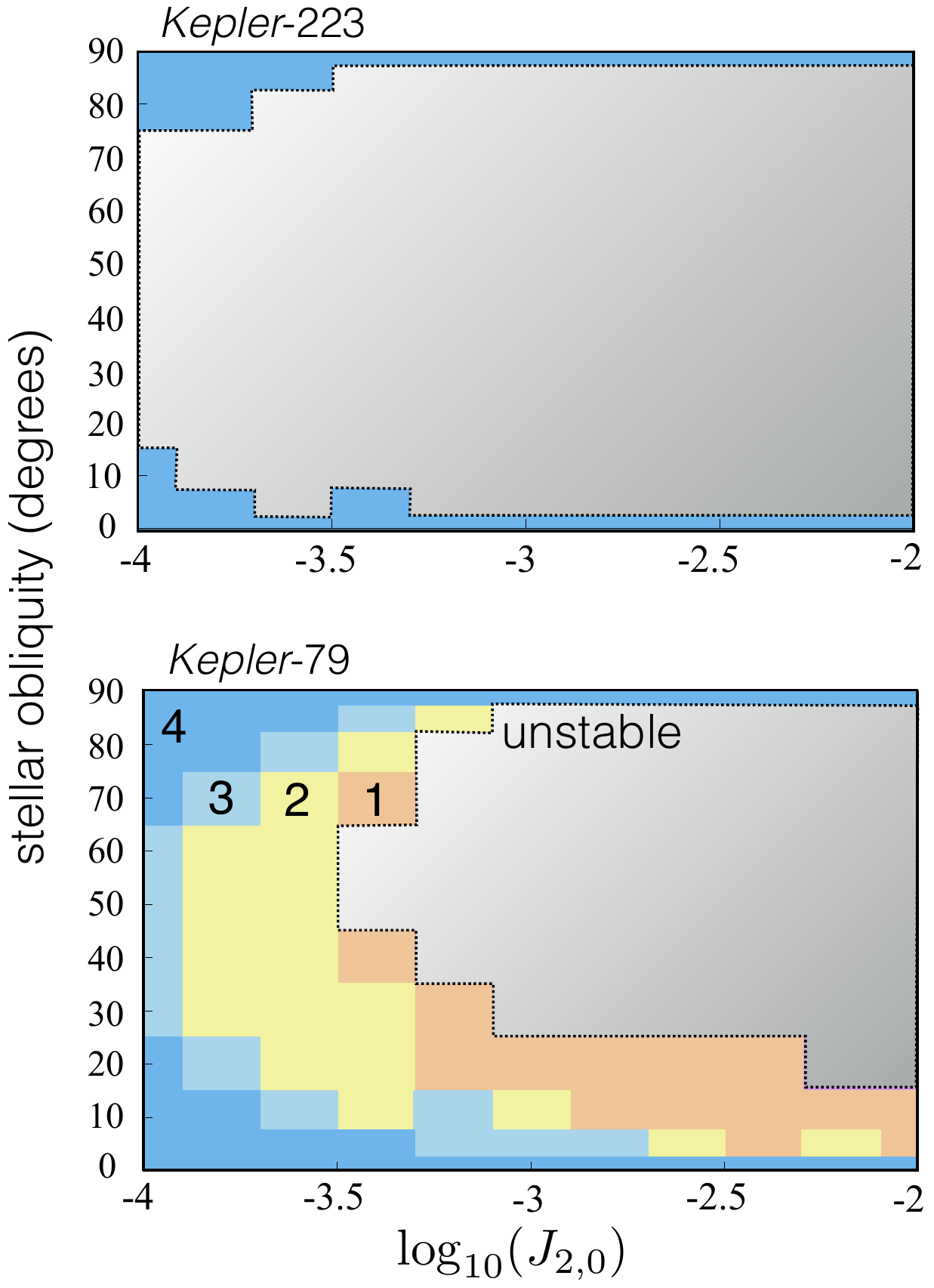}
\caption{The number of planets detectable in transit after 20 million years of simulation from an initially 4-planet configuration. The dotted line outlines the region where one or more planets were lost owing to dynamical instability.}
\label{Results_3}
\end{figure}

For each planetary system, we construct a grid with each cell representing one of the 110 chosen combinations of stellar obliquity and initial $J_{2,0}$. In each cell, the color depicts the maximum number of planets observable in transit, as described above. Systems of 2 planets are illustrated in Figure~\ref{Results_1} and those with 3 or 4 planets are depicted in Figures~\ref{Results_2} \& \ref{Results_3}. The number of co-transiting planets associated with each color is labeled on the figures. 

Crucially, we outline the cases where instability occurred with a dotted line and grey shading. Here, instability is defined as the loss of at least one planet from the system. In reality, the escape velocities of the planets considered are too low to typically remove other planets from the system entirely. Rather, the end result is that planets that are lost will end up either colliding with the star, or colliding with the remaining planets. We do not model the collisions themselves in this work. 


An analytic formula relating the mutual inclination to stellar obliquity and quadrupole moment was presented in \citet{Spalding2016b}\footnote{See their equation (16)}, under the assumptions of circular orbits and low inclinations. For the 2-planet systems in Figure~\ref{Results_1}, we draw a solid black line that denotes this predicted boundary between coplanar and misaligned orbits. The analytic approximations provide a reasonable estimate for the transition between single and double-transiting cases, particularly for $K2-38$ and $Kepler-10$, where the transition occurs at smaller inclinations. 

In order to illustrate the sensitivity of our results to uncertainties in mass measurement, we plot the analytic solution appropriate to the upper and lower bounds of uncertainty upon the planetary masses, as grey lines. In general, uncertainties do not significantly alter the expected demarcation between aligned and misaligned systems, and the real systems are approximately equally likely to be more or less stable. However, the largest region of instability is for \textit{Kepler}-131, making this system particularly poorly suited to a discussion of its own specific history. Nevertheless its masses remain representative of \textit{Kepler} systems in general and so its response to the stellar potential constituents a relevant result.

As stated above, our primary goal was to delineate the ubiquity of stellar oblateness as an instability mechanism. To that end, we note that only \textit{Kepler}-10 was immune to instability for all chosen parameters, with \textit{Kepler}-36 remaining stable all but two times. All other systems were susceptible, at least for the upper range of $J_{2,0}$. Accordingly, we conclude that the instability mechanism described in \citet{Spalding2016b} constitutes a viable pathway toward instability for low and high-multiplicity systems alike. In general terms, the range of $J_{2,0}$ leading to instability is slightly smaller for the 3 and 4 planet systems compared to 2-planet systems, however, given our small sample size such a pattern is by no means statistically significant. 

\subsection{Eccentricities}

If a single-transiting system is observed, it is difficult to infer whether there exist any non-transiting companions. Within the framework of our present investigation, a key outcome of dynamical instability is the presence of eccentricity within the planetary orbits that subsequently remain in the system. Accordingly, eccentricity within a single transiting planet's orbit stands as a detectable signature of the loss of non-transiting companions. However, for the shortest-period systems, tidal effects are likely to have damped out any traces of primordial eccentricity. The tidal circularization timescale is given by \citep{Murray1999}
\begin{align}
\tau_{e}&\equiv \bigg|\frac{e}{\dot{e}}\bigg|\approx\frac{2}{21}\frac{Q}{k_{2,p} n_p}\frac{m_p}{M_\star}\bigg(\frac{a_p}{R_p}\bigg)^5\nonumber\\
&\approx40\bigg(\frac{a_p}{0.1\textrm{AU}}\bigg)^{\frac{13}{2}}\bigg(\frac{Q/k_{2,p}}{1000}\bigg)\bigg(\frac{2 R_{\oplus}}{R_p}\bigg)^5\bigg(\frac{m_p}{10M_\oplus}\bigg)\textrm{Gyr}
\end{align}
where $k_{2,p}$ is the planetary Love number, $Q_p$ is its tidal quality factor, $R_\oplus$ is an Earth radius and $M_\oplus$ is an Earth mass. Put another way, planets possessing 10 Earth masses and 2 Earth radii will circularize within a Gyr for semi-major axes below $a_p\sim0.05$\,AU. Those with semi-major axes exceeding $a_p\sim0.1$\,AU, however, ought to possess eccentricities that are relatively unaffected by tides.

With the caveat regarding tidal circularization in mind, it is interesting to tabulate the orbital parameters of the planet that remains after dynamical instability within the four most unstable 2-planet examples -- $K2$-38, $Kepler$-27, $Kepler$-131 and \textit{Kepler}-307. As can be seen from Table~\ref{Ecc}, the mean eccentricity of the remaining planet is roughly $\bar{e}_{i}\approx0.3-0.4$.

An additional factor worth mentioning is that we tabulated eccentricities from the mercury6 $N$-body code. However, we did not model collisions between planets, which is likely to influence the final eccentricity distribution. Accordingly, the eccentricities in reality may be smaller than we predict here owing to dissipative processes associated with the physics of merging, along with dynamical friction from the production of the associated debris.

Cumulatively, we may propose the following observational signature. First, consider a sample of single-transiting systems beyond 0.1\,AU. Suppose that they are comprised of two populations, a fraction $f_{\textrm{in}}$ that have undergone dynamical instability and a fraction $1-f_{\textrm{in}}$ that have not (for now, the source of instability is left undetermined). The latter fraction did not encounter a dynamical instability, and therefore appear single owing to large mutual inclinations with unseen companions, or alternatively were born single. 

If we now suppose that the unstable population were predominately generated by stellar oblateness and obliquity, they should possess a mean eccentricity of $\bar{e}_{\textrm{inst}}\approx\bar{e}_{i}$. Denoting the mean across both populations as $\bar{e}$ and the mean of the stable population as $\bar{e}_{\textrm{st}}\ll\bar{e}_{\textrm{inst}}$, one can show that
\begin{align}
f_{\textrm{in}}&=\frac{\bar{e}-\bar{e}_{\textrm{st}}}{\bar{e}_{\textrm{inst}}-\bar{e}_{\textrm{st}}}\nonumber\\
&\approx\frac{\bar{e}}{\bar{e}_{\textrm{inst}}},
\end{align}
where the second equality assumes the stable population will exhibit eccentricities much lower than the unstable population.

Typically, the \textit{Kepler} Dichotomy is quoted as reflecting a roughly equal split between the large and small inclination systems, i.e., $f_{in}=1/2$ \citep{Johansen2012,Ballard2016}. In order to reproduce this fraction with $\bar{e}_{u,o}\approx0.4$, we would predict $\bar{e}_o\approx0.2$. There are of course numerous other dynamical interactions capable of exciting, or indeed damping, eccentricities. Furthermore, separate pathways to instability exist that will produce their own eccentric population of planets. These include planet-planet scattering or the presence of an external perturber, as mentioned above, among others \citep{Obrien2006,Ford2008,Lai2017,Becker2017,Mustill2017,Hansen2017}.

In order to deduce the dominant driver of instabilities, it is essential to determine the expected eccentricity distribution of each mechanism, and their their typical occurrence rate. For example, the occurrence rate of instability driven by an exterior companion is limited by the abundance of exterior companions, which is an active area of research (e.g., \citealt{Wang2015,Bryan2016}). Until the statistics of these other mechanisms are investigated in detail, the ultimate source of instability shall be difficult to decipher. Nonetheless the above discussion outlines the feasibility of deriving the true underlying abundance of planets despite only observing the proportion that transit. 

\subsection{Semi-major axes}
During the planets' close encounters with each other at high eccentricity, the semi-major axes of both planets are altered. The planet that remains usually ends up with an increased semi-major axes, whereas the other planet typically collides with the star. To that end, recall that the stellar radius was held fixed at a larger value in the simulations, and $J_2$ was forced to decay. This prescription is correct in terms of the star's gravitational influence. In real systems, however, the star would have contracted somewhat by the time instability occurs, and so more energy would need to be transferred to collide with the smaller star. 

As a consequence of the details outlined above, larger semi-major axes and/or eccentricity alterations might occur in reality than we see in our simulations. Additionally, tides may ``save" the inner planet during a high-eccentricity phase, by damping its eccentricity before it enters a star-crossing trajectory (generating so-called ``ultra short period planets"; \citealt{Adams2016}). These details of the problem do not alter the general picture, but will influence the statistical properties of any proposed population of post-instability planets.

Though the quantitative nature of our predictions are subject to numerous uncertainties, the qualitative prediction is that a population of single-transiting systems ought to exhibit larger eccentricities than those possessing unseen companion planets.

\section{Mechanism of instability}

If the stellar quadrupole only induced instability in systems with 3 or more planets, it would have been difficult to understand, in simple terms, the physical mechanism behind it. However, the onset of instability in 2-planet systems leaves the process amenable to semi-analytic investigation, in order to attain a deeper understanding. In this section, we explore the problem from such an analytic point of view, using $K2-38$ as a test case.
\subsection{Analytic Treatment}
 In \citet{Spalding2016b}, 2-planet systems were studied analytically by expanding the gravitational interaction potential between the two planets to lowest (second) order in inclinations, with eccentricities fixed to be zero (Lagrange-Laplace secular theory; \citealt{Murray1999}). This approach yielded a closed-form solution for the relative inclination excited between the two planetary orbits. The locus of stellar $J_2$ and $\beta_\star$ that takes the two transits out of the same plane is drawn onto Figure~\ref{Results_1}, and agrees relatively well with the transition between coplanar and misaligned systems. Despite this approximate agreement, the Lagrange-Laplace framework is ill-equipped to explain why greater inclinations or oblateness give rise to instability (in part owing to the decoupling of eccentricity and inclination dynamics to second order).
 
 In order to study the onset of instability, we lift the assumptions of circular orbits and low inclination, by utilising an expansion of the disturbing potential that uses the semi-major axis ratio as a small parameter \citep{Kaula1962}:
\begin{align}\label{KaulaHam}
\mathcal{H}=&\frac{G m_1 m_2}{a_2}\sum_{l=2}^{l=\infty}\bigg(\frac{a_1}{a_2}\bigg)^l A_l(e_1,e_2,I_1,I_2)\nonumber\\
&\times \cos(j_1\lambda_1+j_2\lambda_2+j_3\varpi_1+j_4\varpi_2+j_5\Omega_1+j_6\Omega_2).
\end{align}
Note that the above expansion is written in a reduced form, with significant information encoded in the value of $A_l$. In particular, each order of $l$ possess numerous terms with different cosine arguments and pre-factors. We will only include terms of orders $l=2,3$ and $4$, referred to as quadrupole, octupole and hexadecapole respectively (see below). The constants $j_i$ are constructed such that $\sum_{i=1}^6j_i=0$ \citep{Murray1999}, and the angles $\lambda_i$, $\varpi_i$ and $\Omega_i$ are respectively the mean longitude, longitude of pericenter and longitude of ascending nodes of the planetary orbits.

 \begin{table*}
  \centering
\begin{tabular}{ |p{2cm}||p{1.4cm}|p{2cm}|p{1.4cm}|p{2cm}|p{1.2cm}|p{1.2cm}|p{1cm}|  }
 \hline
 \multicolumn{8}{|c|}{2-planet systems} \\
 \hline
System & $a_{1,i}\,(AU)$ & $\bar{a}_{1,f}$\,(AU) & $a_{2,i}\,(AU)$ & $\bar{a}_{2,f}$\,(AU)  & $\bar{e}_{1,f}$& $\bar{e}_{2,f}$& $\bar{e}_{f}$\\
 \hline
 \textit{K2}-38 & 0.0505 & 0.0732 &0.0965& 0.1815 & 0.4507 & 0.4091& 0.4242\\
 \textit{Kepler}-29 & 0.0922 & 0.1502& 0.1090 & 0.1418 & 0.4028 & 0.3351 & 0.3701\\
 \textit{Kepler}-131 & 0.1256& 0.1488 & 0.1708 & 0.2541$^{**}$  & 0.3572 & 0.3840$^{**}$ &0.3603\\
 \textit{Kepler}-307 & 0.0904 & 0.0952 & 0.105 &  0.1378$^*$  & 0.3195 &0.3665$^*$&0.3224\\
 \hline
\end{tabular}  
 \caption{The semi-major axes and eccentricities of the 4 most unstable 2-planet systems resulting from our simulations. For each case where instability occurred, we recording the eccentricity and semi-major axis of the remaining planet, then took the mean of all the results (denoted by an overbar, with the subscript `f' meaning `final,' `i' representing `initial" and the number corresponding to the particular planet). The mean is only a very general guideline as to what to expect, but the results suggest that a population of single-transiting systems that had undergone our proposed instability mechanism would be expected to yield an average eccentricity of roughly 0.3-0.4.}
  \label{Ecc}
\end{table*}
The above Hamiltonian contains infinite ``harmonics" -- the cosine terms -- each associated with its own specific resonance. Here, a resonance may be thought of as a restoring torque that tends to force libration about some constant value of the argument. If we can assume that the system is close to one of these resonances, and no other resonances overlap the associated phase-spaced domains, it is possible to ignore the other harmonics and consider the dynamics associated with one harmonic alone \citep{Lichtenberg1992,Morby2002}. In order to determine which harmonic(s) drive the observed dynamics, we turn to our numerical simulations. 
\begin{figure}[h!]
\centering
\includegraphics[trim=0cm 0cm 0cm 0cm, clip=true,width=1\columnwidth]{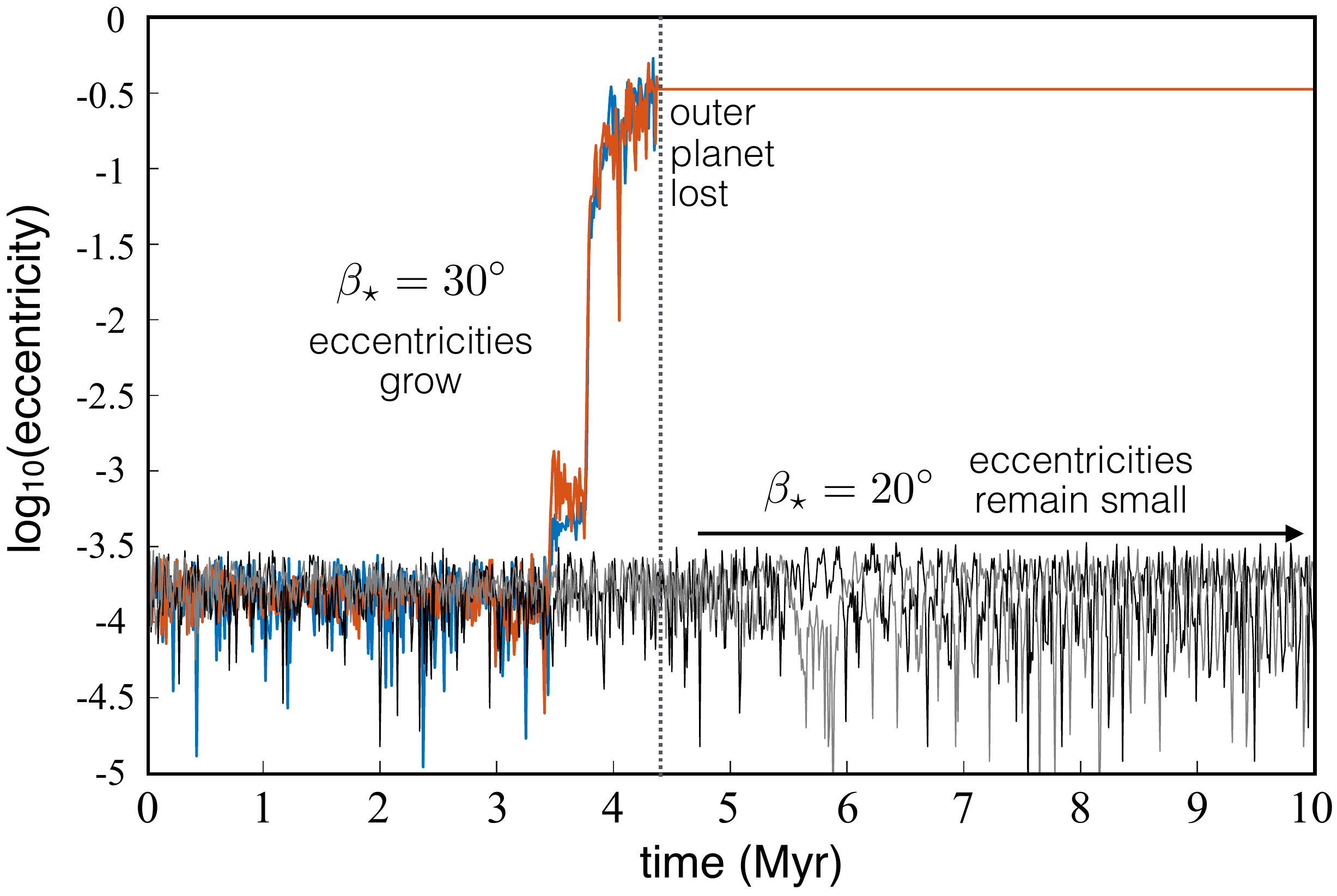}
\caption{The evolution of eccentricity of both planets in the $K2$-38 system when the stellar obliquity is set at $30^\circ$ (blue, inner planet and red, outer planet) and $20^{\circ}$ (grey, inner planet and black, outer planet). For both cases, oblateness decays from $J_{2,0}=10^{-2.6}$. The difference in dynamics between the two cases is profound. Whereas at $20^\circ$ both planets remain circular, at $30^\circ$ both eccentricities begin to grow in unison at 3.75\,Myr. After reaching eccentricities of roughly 1/2, instability sends the outer planet into the central star. We discuss this process in the text.} 
\label{Ecc1}
\end{figure}
\begin{figure}[h!]
\centering
\includegraphics[trim=0cm 0cm 0cm 0cm, clip=true,width=1\columnwidth]{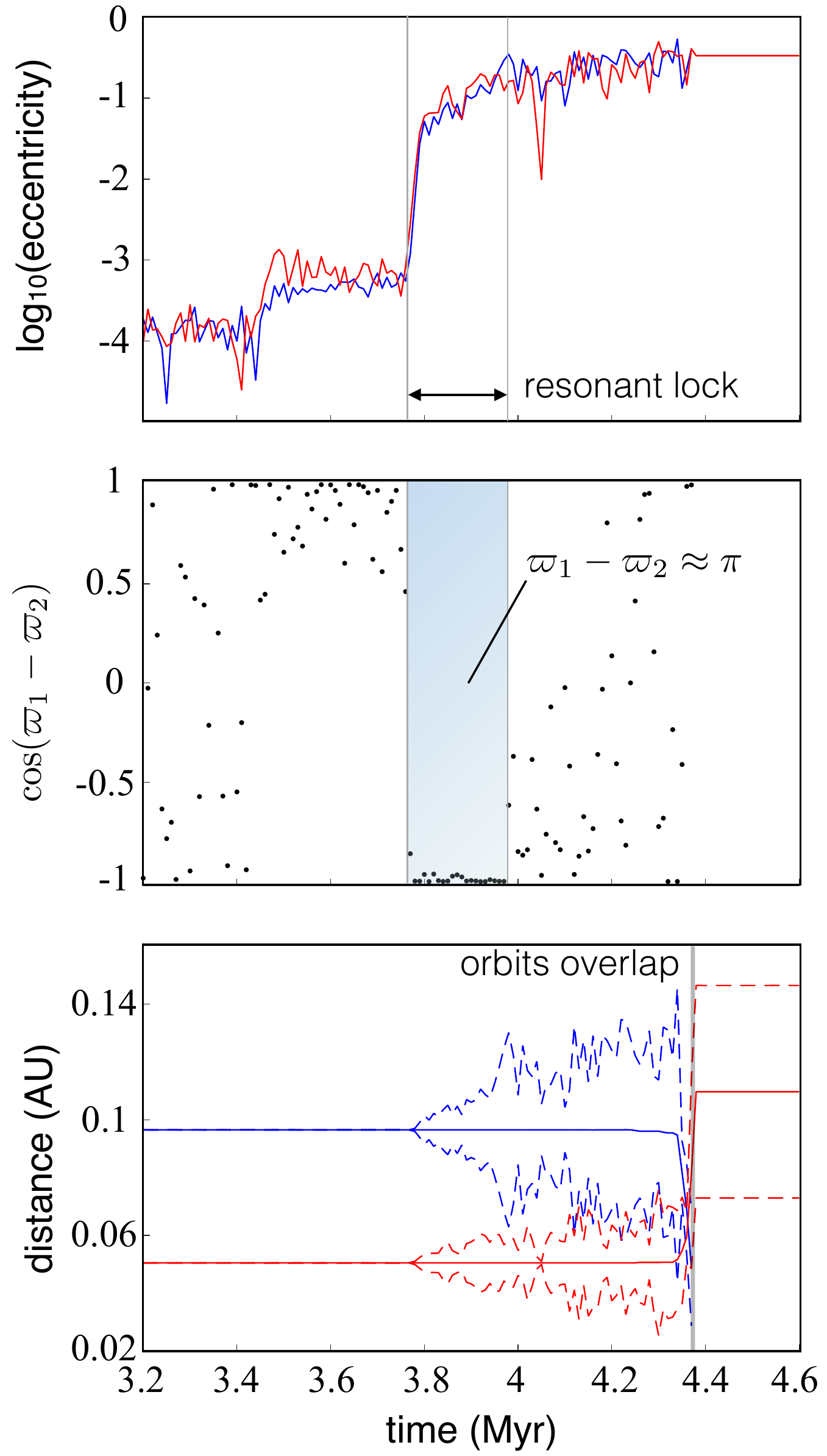}
\caption{An illustration of dynamics close to the time of instability of $K2$-38 with parameters $\beta_\star=30^\circ$, $J_{2,0}=10^{-2.6}$. Top panel: The evolution of eccentricity as a function of time for the outer planet (blue) and the inner planet (red). Middle panel: Time evolution of the resonant argument $\cos(\varpi_1-\varpi_2)$ through instability. Notice that the argument librates close to $\pi$ during the main phase of eccentricity growth (the shaded, blue region), which is indicative of secular resonant capture. During these dynamics, $\dot{\varpi}_1\approx\dot{\varpi}_2$. Bottom panel: Illustration of ultimate cause of instability. The solid lines illustrate semi-major axis of the inner (red) and outer (blue) planets, whilst the dotted lines denote the apocenter (upper) and pericenter (lower) of the orbits. Secular resonance is broken as the orbits begin to cross (time\,$\approx3.98$\,Myr), and instability ensues soon after (time\,$\approx4.36$\,Myr). } 
\label{resonance}
\end{figure}

Looking specifically at $K2$-38, we consider two cases, one that undergoes instability and another that does not, annotated in Figure~\ref{Results_1}. The same value of $J_{2,0}=10^{-2.6}$ is chosen, with the stable case having $\beta_\star=20^{\circ}$ and the unstable case corresponding to $\beta_\star=30^{\circ}$. In Figure~\ref{Ecc1} we plot the eccentricities of both planets as a function of time at each value of stellar obliquity. Notably, in the stable 20$^\circ$ case, both eccentricities remain low, but the dynamics change qualitatively at $30^\circ$. In this latter, unstable case, both eccentricities begin to grow simultaneously at $\sim3.75$\,Myr until roughly 0.6\,Myr later, when the system undergoes instability and the outer planet is lost through collision with the central star.

Eccentricity growth of the kind described above is a common outcome of capture into a secular resonance \citep{Ward1976,Batygin2016}, whereby two precession frequencies become roughly commensurate, causing them to ``lock" as system parameters evolve. In order to deduce which resonance the system enters we illustrate the evolution of the argument $\varpi_1-\varpi_2$ in Figure~\ref{resonance} (middle panel). Concurrent with the initiation of eccentricity-growth, the system enters a libration of $\varpi_1-\varpi_2$ around $\pi$, i.e., the orbits are roughly anti-aligned. Accordingly, the resonance corresponds to a commensurability between the frequencies $\dot{\varpi}_1$ and $\dot{\varpi}_2$. 

Interestingly, $\varpi_1-\varpi_2$ appears to librate, with a large amplitude, around $\varpi_1-\varpi_2=0$ for a brief period before the resonant growth of eccentricity begins. Furthermore, this brief period of apparent libration corresponds to an order of magnitude increase in eccentricity, from $e_i\sim 10^{-4}$ to $e_i\sim 10^{-3}$. This libration does not imply resonant locking. A circulating trajectory in phase-space will appear to librate if the center of libration is offset from the origin and the libration amplitude is small enough \citep{Lichtenberg1992,Morby2002}.

With the understanding that eccentricity growth commences at $\dot{\varpi}_1\approx\dot{\varpi}_2$, we can begin to develop a criterion for the onset of instability within a given planetary system. We expand Hamiltonian~\ref{KaulaHam}, but remove all harmonics except for $\cos(\varpi_1-\varpi_2)$. For illustration, we expand the potential to fourth (hexadecapolar) order, but in order to treat the secular resonant dynamics at high precision, higher order expansions are likely required (e.g., \citealt{Boue2012}). The disturbing function acting between the two planets may then be written as: 
\begin{align}
\mathcal{R}_{12}&=\mathcal{R}_{\textrm{quad}}+\mathcal{R}_{\textrm{oct}}+\mathcal{R}_{\textrm{hexa}}\nonumber\\
\mathcal{R}_{\textrm{quad}}&=\frac{G m_1m_2}{a_2}\bigg(\frac{a_1}{a_2}\bigg)^2\frac{2+3 e_1^2}{128(1-e_2^2)^{3/2}}\nonumber\\
&\times\big[1+3\cos(2 I_1)\big]\big[1+3\cos(2 I_2)\big]\nonumber\\
\mathcal{R}_{\textrm{oct}}&=-\frac{G m_1m_2}{a_2}\bigg(\frac{a_1}{a_2}\bigg)^3\frac{15e_1e_2(4+3e_1^2)}{4096(1-e_2^2)^{5/2}}\cos(\varpi_1-\varpi_2)\nonumber\\
&\times\big[5\cos(I_1)\big(3\cos(I_1)-2\big)-1\big]\big[(1+\cos[I_1])\big]\nonumber\\
&\times\big[5\cos(I_2)\big(3\cos(I_2)-2\big)-1\big]\big[(1+\cos[I_2])\big]\nonumber\\
\mathcal{R}_{\textrm{hexa}}&=\frac{G m_1m_2}{a_2}\bigg(\frac{a_1}{a_2}\bigg)^4\Bigg[\frac{9(15e_1^4+40e_1^2+8)(3e_2^2+2)}{4194304(1-e_2^2)^{7/2}}\Bigg]\nonumber\\
&\times \big(20 \cos (2 I_1 )+35 \cos (4 I_1 )+9\big)\nonumber\\
&\times \big(20 \cos (2 I_2 )+35 \cos (4 I_2 )+9\big).
\end{align}
In addition to the planet-planet disturbing potential, the stellar disturbing potential may be written as \citep{Danby1992}
\begin{align}\label{starDisturb}
\mathcal{R}_{J_2,p}=-\frac{G m_pM_\star}{2a_p}J_2\bigg(\frac{R_\star}{a_p}\bigg)^2\bigg[\frac{3}{2}\sin^2(I_p)-1\bigg](1-e_p^2)^{-\frac{3}{2}}.
\end{align}

In order to solve for the inclinations at which $\dot{\varpi_1}\approx\dot{\varpi}_2$, we use Lagrange's planetary equations\footnote{Note that we choose a form for the disturbing function that yields units of energy. Notation elsewhere does not include a factor of $m_p$  where here $p$ refers to the planet experiencing a perturbation. Accordingly, in equation (18) we must include an extra factor of $m_p$ on the left hand side.} \citep{Murray1999}
\begin{align}\label{equ}
m_p\sqrt{GMa_p}\frac{d\varpi_p}{dt}=&\frac{\sqrt{1-e_p^2}}{e_p}\frac{\partial \mathcal{R}}{\partial e_p}+\frac{\tan(I_p/2)}{\sqrt{1-e_p^2}} \frac{\partial\mathcal{R}}{\partial I_p}\nonumber\\
&-m_p\sqrt{GMa_p}\frac{3 G M_\star }{c^2a(1-e_p)}n_p
\end{align}
where the full disturbing function is given by
\begin{align}\label{DisturbingVarpi}
\mathcal{R}=\mathcal{R}_{J_2,1}+\mathcal{R}_{J_2,2}+\mathcal{R}_{12},
\end{align}
and we have introduced the speed of light $c$ through the inclusion of general relativistic precession \citep{Wald2010}. 

\subsection{Onset of secular resonance}
\begin{figure*}
\centering
\includegraphics[trim=0cm 0cm 0cm 0cm, clip=true,width=1\textwidth]{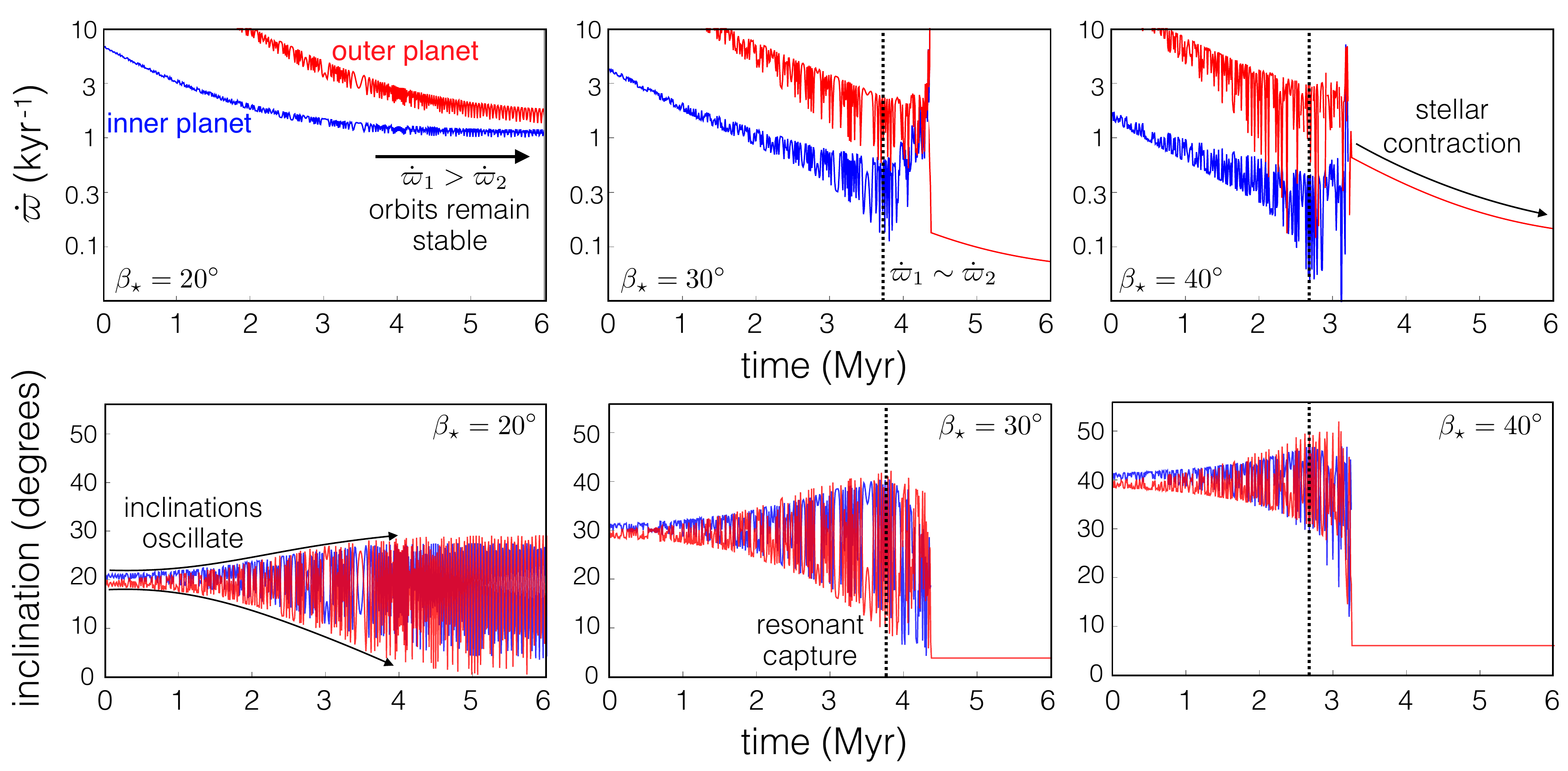}
\caption{The evolution of the precession rate of the longitude of pericenter ($\dot{\varpi}$, top panel) and the inclinations (bottom panel), for both planets orbiting $K2-38$. All cases have $J_{2,0}=10^{-2.6}$ and we plot 3 cases, including the stable $\beta_\star=20^\circ$ (left) and two unstable cases, $\beta_\star=\{30^\circ,\,40^\circ\}$. In all cases, the inner planet is represented by blue and the outer planet by red. To construct the evolution of $\dot{\varpi}$, we inserted the orbital parameters (inclinations, eccentricties and semi-major axes) computed from our simulations into the expression acquired from equation~\ref{equ}. In addition, we averaged the $\varpi_1-\varpi_2$ angle as described in the text. The vertical dotted line denotes the point at which eccentricities begin to grow, indicating secular resonant capture. Resonance generally occurs when $\dot{\varpi}_1\sim\dot{\varpi}_2$.} 
\label{Inc_Evolution}
\end{figure*}

The precession frequencies $\dot{\varpi}_p$ depend upon the orbital inclinations $I_1$ and $I_2$. In Figure~\ref{Inc_Evolution}, we plot the evolution of both planetary inclinations and precession frequencies $\dot{\varpi}_p$ in the $K2-38$ system as a function of time. We choose to illustrate the two cases discussed above, with $J_{2,0}=10^{-2.6}$ and $\beta_\star=\{20^\circ,\,30^\circ\}$, but include a third, unstable case, with $\beta_\star=40^\circ$. 

In order to construct Figure~\ref{Inc_Evolution}, we carry out the differentiation presented in equation~\ref{equ} in order to obtain a closed-form expression for $\dot{\varpi}_p$ as a function of the planetary orbital parameters. We then inserted the orbital parameters (inclinations, eccentricties and semi-major axes) as they emerge from our simulations into this expression, together with $R_\star=2R_\odot$ and $J_2$ as given by equation~\ref{J2}. However, given that outside of resonance the argument $\varpi_1-\varpi_2$ circulates on a relatively short timescale, we averaged over this harmonic (equivalent to setting $\varpi_1-\varpi_2=\pi/2$) in order to illustrate the dynamics that lead to secular resonant capture.

Inclinations plotted in Figure~\ref{Inc_Evolution} are obtained directly from the simulations. Both inclinations begin equal to $\beta_\star$, the stellar obliquity, but begin to oscillate with ever-increasing amplitude as the stellar quadrupole decays. As the inclinations evolve, both planetary precession frequencies trend toward lower values, which is primarily a consequence of the stellar quadrupole weakening. For $\beta_\star=20^\circ$, the inclinations both remain below $\sim30^\circ$, and $\dot{\varpi}_1$ remains greater than $\dot{\varpi}_2$ for the duration of the simulation. Consequently, secular resonance is not encountered.

In the higher-obliquity cases, the inclinations oscillate sufficiently widely to lead to a situation when $\dot{\varpi}_1\sim\dot{\varpi}_2$. The vertical dotted lines indicate when eccentricity growth begins, which corresponds well with the time at which $\dot{\varpi}_1\sim\dot{\varpi}_2$ for $\beta_\star=30^\circ$, though only approximately in the case where $\beta_\star=40^\circ$.

We illustrate the above discussion in a different form using Figure~\ref{Locus}. Here, we again look at $K2-38$ and $J_{2,0}=10^{-2.6}$. However, the colored contours denote the locus of $I_1$ and $I_2$ at which $\dot{\varpi}_1\sim\dot{\varpi}_2$ at various times within the simulation. The contours move as the stellar quadrupole weakens. We plot blue points to represent the 10 initial conditions upon inclination used in our simulations. From these inclination configurations, as can be seen from Figure~\ref{Inc_Evolution}, the inclinations of both planets oscillate, but they do so in an anti-correlated fashion such that their trajectory in Figure~\ref{Locus} follows an arc, as illustrated schematically. As this trajectory crosses the lines of $\dot{\varpi}_1\sim\dot{\varpi}_2$, in a broad sense, secular resonant may be encountered.

This expectation that secular growth of eccentricity coincides with $\dot{\varpi}_1\sim\dot{\varpi}_2$ is only loosely in agreement with Figure~\ref{Inc_Evolution}. The reason for the imperfect agreement most likely arises because the locus of $\dot{\varpi}_1-\dot{\varpi}_2\approx0$ is a measure of the instantaneous precession frequencies. However, in reality the inclinations are changing with time on a similar timescale to libration in the eccentricity degree of freedom. Accordingly, a more rigorous treatment, taking account of secular resonant capture within a 2 degree of freedom framework, is required to improve upon the current description. Despite the resonant criterion failing quantitatively, the qualitative picture remains unchanged.



\subsection{Requirement of large inclinations}

\begin{figure}
\centering
\includegraphics[trim=0cm 0cm 0cm 0cm, clip=true,width=1\columnwidth]{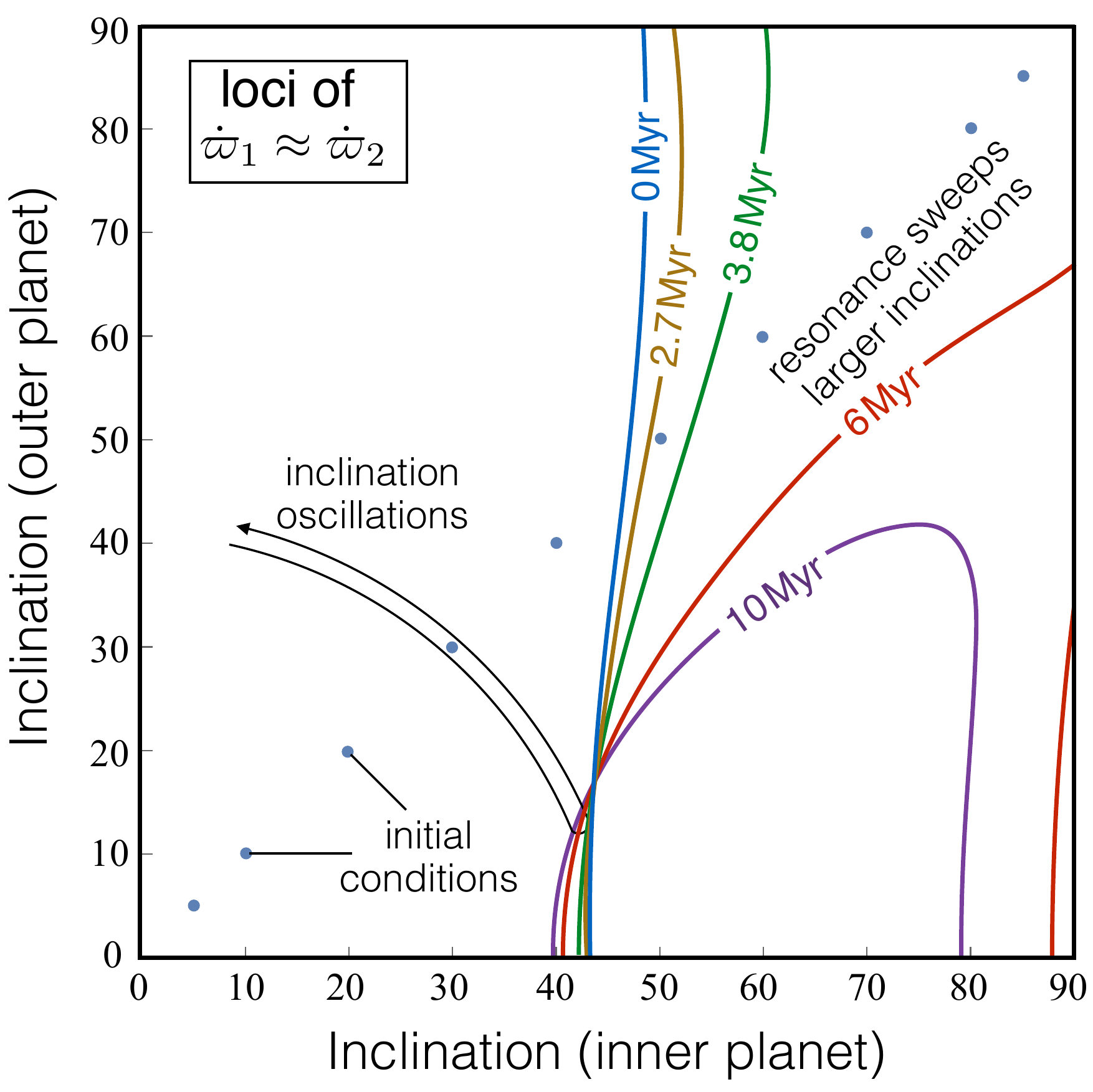}
\caption{Contours of $\dot{\varpi}_1\sim\dot{\varpi}_2$ at different times, beginning with $J_{2,0}=10^{-2.6}$. Times plotted are 0\,Myr (blue), 2.7\,Myr (yellow, approximately the time of instability for $\beta_\star=40^\circ$), 3.8\,Myr (green, approximately the time of instability for $\beta_\star=40^\circ$), 6\,Myr (red) and 10\,Myr (purple). All 10 initial conditions are plotted as blue points, i.e., the 10 values of stellar obliquity modelled. From each blue point, the trajectories of each simulation begin oscillations in both inclinations (as in Figure~\ref{Inc_Evolution}) until they intersect the appropriate curve of $\dot{\varpi}_1\sim\dot{\varpi}_2$. At approximately that point, secular resonance is enountered and instability soon develops. It should be noted that for stellar obliquities above $\beta_\star\sim50^\circ$, the curve of $\dot{\varpi}_1\sim\dot{\varpi}_2$ will sweep past the initial inclinations and typically cause instability even without inclination oscillations.} 
\label{Locus}
\end{figure}

We emphasise that the above expressions do not make any assumptions regarding inclinations. This aspect is key, because at small inclinations no configuration exists that brings the two precession frequencies close to one another (see contours in Figure~\ref{Locus}). However Figure~\ref{Locus} indicates that when the inner planet is inclined by more than $\sim$40\,$^\circ$, the two frequencies can be brought close to one another. 

The requirement of planetary inclinations may be understood by noting that the inner planet's greater proximity to the star contributes to a faster $J_2$-induced precession rate in the coplanar case. However, as the inner planet is tilted, the stellar quadrupole's influence weakens, such that there exists a critical inclination at which the two planets are precessing at equal rates. Though different in important aspects, the effect whereby higher inclinations open up a system to secular resonant behaviour is reminiscent of the Kozai-Lidov resonance, which has found wide-spread usage within celestial mechanics \citep{Lidov1962,Kozai1962,Fabrycky2007,Nagasawa2008,Naoz2011}. The resonance we outline may likewise have had wide-spread importance in the evolution of systems around oblate central bodies. 

Secular resonances do not exist at low inclinations in $K2$-38 owing to the low angular momentum of the inner body relative to the outer body. The planet-planet induced precession cannot overcome the greater influence of the stellar quadrupole at shorter orbital periods. It was found in \citet{Spalding2016b} that resonance in the argument of ascending node only existed if the inner planet possessed more angular momentum than the outer planet. A similar scenario is found here. It is possible to find low-inclination resonant values of $J_2$ in the systems, such as $Kepler$-131 that possess an inner planet with more angular momentum than the outer planet. However, the resonant value of $J_2$ is an order of magnitude larger than the largest value we considered and thus plays no role in these dynamics. High inclinations must be excited if the system is to enter resonance.

To close our discussion of the instability itself, we illustrate physically why eccentricity growth leads to instability. In the bottom panel of Figure~\ref{resonance} we plot the pericenters, apocenters and semi-major axes of both planets in the unstable case. Instability corresponds roughly to the time when the pericenter of the outer planet coincides with the apocenter of the inner planet. If the orbits were perfectly anti-aligned and in the same plane this configuration corresponds to orbit-crossing. Whereas they are not in the same plane in general, their libration around $\varpi_1-\varpi_2=\pi$ suggests the orbits come close to crossing.

\section{Conclusions}

\subsection{Ubiquity of instability}

The primary motivation for this work was to determine whether the gravitational perturbation arising from a tilted, oblate star is sufficient to destabilize systems of planets possessing low multiplicity. We studied 11 systems, 6 of which possess 2 planets, 3 possess 3 planets and 2 possess 4 planets. We find that instability occurred in all but one system (\textit{Kepler}-10), though in general instability only occurred for $J_2\gtrsim10^{-3}$ and stellar obliquities $\beta_\star\gtrsim 30^{\circ}$, with the range varying widely (see Figures~\ref{Results_1}, \ref{Results_2} \& \ref{Results_3}). 

Having studied only 11 systems, we are unable to place precise, quantitative constraints at a population level upon the prevalence on instability. However, if we suppose that $J_2\gtrsim10^{-3}$ leads to instability in most systems, as appears generally to be the case in our small sample, then this equates to periods
\begin{align}
P_\star\lesssim 20\pi\bigg(\frac{R_\star^3}{GM_\star}\bigg)^{1/2}\sim3\,\textrm{days},
\end{align}
but the critical value can vary for different assumptions on the appropriate stellar radius. T-Tauri stars spin with periods ranging between about 1-10 days, with the median of the distribution lying close to 3-5 days \citep{Bouvier2013}. Furthermore, there is evidence that stars spin up slightly to periods below 3 days immediately following disk dissipation \citep{Bouvier2014,Karim2016}. These observations suggest that a relatively large fraction, perhaps as many as 1/2 of systems are subject to this instability. 


In addition to the proportion of systems exhibiting large enough quadrupole moments, we must also consider the distribution of stellar obliquities. By inspection of Figures~\ref{Results_1}, \ref{Results_2} \& \ref{Results_3}, instability generally occurs only when stellar obliquity exceeds $\sim30^\circ$, though exceptions exist. The stellar obliquity of hot stars (surface temperature above 6200\,K) hosting hot Jupiters appears to be close to isotropic \citep{Winn2010,Albrecht2012}. Around such objects, if instability was triggered for $150^\circ>\beta_\star>30^\circ$\footnote{We are implicitly assuming that a stellar obliquity of $30^\circ$ is dynamically equivalent to one of $150^\circ$. This will be true of the dynamics are dominated by secular interactions, but may not be true when the planets are close to mean motion resonances, when their mean anomalies become important for the dynamics.}, we would expect an unstable fraction given by
\begin{align}
f_{30}&=\frac{\int_{30}^{150}\sin(\theta)d\theta}{\int_0^{180}\sin(\theta)d\theta}\nonumber\\
&\sim0.9.
\end{align}
This fraction is close to unity, and so naively will not significantly reduce the fraction of 1/2 above for stars that are rapidly-rotating enough to induce instability. However, the picture changes for cool stars and smaller planets \citep{Li2016,Winn2017}, where the obliquities appear substantially reduced (though values up to 30$^\circ$ still occur in these systems). 

Given the requirement of a large obliquity, we consider the fraction of stars spinning fast enough to cause instability, 1/2, as an upper limit, with the underlying primordial distribution of stellar obliquities reducing this fraction by a currently-unknown amount. Depending upon the true values of many uncertain parameters, the instability mechanism might in principle turn out to be almost ubiquitous, or extremely rare. As new generations of observational surveys come online, the origin and abundance of stellar obliquities will come into clearer focus, as will the ubiquity of the instability outlined here.

  In addition to uncertainties, our estimates above are limited by the so-far small sample size of 11 modelled systems, poor knowledge of young star radii and rotational evolution, along with the present dearth of spin-orbit misalignment measurements in systems of lower-mass planets \citep{Wang2017}. With those caveats in mind, the approximate, yet slightly optimistic discussion above suggests that up to 1/2 of super-Earth systems might pass through a phase where their host star's quadrupole moment triggers instability.

\subsection{Observational tests}

An additional goal of this work was to progress toward a method of distinguishing single-transiting systems with unseen transiting companions from those systems possessing a single planet intrinsically. One way to accomplish this directly is through the measurement of transit timing variations arising both from direct perturbations upon the transiting planet  \citep{Agol2005,Nesvorny2012}, and from astrometric variations of the stellar light curve induced from the perturbations upon the star itself \citep{Millholland2016}. However, here we propose that if the stellar oblateness drives instability in a significant fraction of systems, one may distinguish single transiting from single planet systems at a population level by measuring the eccentricities of the transiting planets. We find that typical eccentricities excited lie between 0.3 and 0.4 (Table~\ref{Ecc}), and that tidal circularization is ineffective at erasing these eccentricities provided the planet resides outside of roughly 0.1\,AU.

Given that the stellar quadrupole falls of as the square of semi-major axis, we would expect that the mechanism is less effective for more distant systems. Indeed, in a general sense, we would predict that the closest single-transiting planets exhibit low eccentricities, owing to tides. A little further away we would expect the eccentricities to grow, before decaying again as the instability mechanism becomes less effective. Uncertainties on tidal dissipation, together with the influence of semi-major axis upon stellar obliquities \citep{Li2016,Dai2017} make a prediction for the value of the proposed peak somewhat speculative.

\subsection{Constraints upon stellar obliquity}

Our analysis of planetary system stability allows us to place loose upper bounds upon stellar obliquity in order for specific multi-planet systems to have remained coplanar. For example, we predict that the stellar obliquity of $K2$-38 is under $\sim$20$^\circ$, otherwise the two planetary orbits ought to have been misaligned with one another. Likewise $Kepler$-10 is probably no more misaligned with its planetary orbits than $\sim 20^{\circ}$. 

We are hesitant to make similar predictions regarding $Kepler$-223, as although it appeared highly unstable in our integrations, we did not take care to reproduce the multi-resonant configuration as is currently observed \citep{Mills2016}, which might help retain the planets within the same plane. Indeed, it is interesting to note that $Kepler$-223 and $Kepler$-60 were the most unstable systems in our sample. The former is known to exhibit a 4-body resonance, and the latter may or may not be within such a configuration \citep{Jontof-Hutter2016}. Future work would benefit from analysing the ability for mean motion resonances to ``protect" planetary systems from instability mechanisms such as the stellar quadrupole.

\subsection{Future considerations}

This work considered an initial condition whereby the planetary orbits were coplanar, assuming the disk to have dissipated more rapidly that the orbits can reconfigure into their equilibrium potential. Future treatments should consider this aspect. In particular, the disk itself leads to a precession of longitudes of periapse for embedded planets. Given our finding that the instability is driven by a resonance between $\dot{\varpi}$ of planetary pairs, it would be a fruitful investigation to consider how the disk's gradual dissipation alters the secular phase space \citep{Ward1981}.  

We obtained a qualitative understanding of the instability mechanism, namely, that the values of $\dot{\varpi}$ of both planets can be brought close to together by way of quadrupole-driven inclinations. The resulting resonance leads to eccentricity growth and orbit-crossing. However, we did not treat this problem in a full, 2-degree of freedom framework. To do so would better elucidate the quantitive criteria governing the instability. Furthermore, within this framework the disk potential may be added as an additional term, and the stellar orientation may be allowed to evolve with time, providing an analytic framework for following the system all the way from formation within a massive disk, to the onset of instability subsequent to disk dispersal.

Here, we used only the best-fit masses and began from zero eccentricities. Two directions of future work would benefit from a more statistical approach, such as Markov-chain Monte Carlo, where initial conditions are drawn from a probability distribution. The first is if one truly sought to reconstruct the previous history of a given system, one would simulate a selection of masses drawn from the observational errors. The second would be to extract masses from a probability distribution that incorporates all measured planetary systems, in order to analyse the mechanisms on a global scale, i.e, to construct fictitious systems, but those whose properties are informed by the real population. Our approach is somewhat intermediate, essentially to test whether the instability might be important. Our results indicate that it is potentially important for many planetary systems and is worth further investigation.

\subsection{Closing Remarks}
Cumulatively, we have shown that the contraction of the host star, an evolutionary phase common to all planetary systems, can play a key role in sculpting the resulting planetary systems. Our own solar system was likely not sensitive to the Sun's quadrupole moment owing to the relatively large semi-major axis of Mercury. Its enhanced stability is in part responsible for Earth's low eccentricity and stable conditions over billions of years. By turning toward exoplanetary systems, we see that the host star is not always the giver of life that is in our system, but rather its gravity may disrupt and destroy the tranquility of the systems it hosts.

\begin{acknowledgments}
 This research is based in part upon work supported by NSF grant AST 1517936 and the NESSF Graduate Fellowship in Earth and Planetary Sciences (C.S). KS thanks the David and Lucile Packard Foundation for their generous support. We thank Erik Petigura, Josh Winn and Greg Laughlin for insightful conversations, and the referee, Juliette Becker, for an insightful report that led to substantial improvements in the manuscript. 
\end{acknowledgments}

\end{document}